\documentclass[10pt,oneside]{amsart}

\usepackage{amstext,amsmath,amsfonts,amscd,amsthm}
\usepackage{cite}

\theoremstyle{plain}

\theoremstyle{definition}
\newtheorem{definition}{Definition}

\newtheorem*{smoothness_principle}{Smoothness Principle}

\theoremstyle{remark}
\newtheorem{remark}{Remark}
\newtheorem{note}{Note}

\begin{document}

\title[]{Spectral geometry of spacetime${ }^{1}$}\thanks{${ }^{1}$
Talk presented at the Euroconference on
      "NON-COMMUTATIVE GEOMETRY AND HOPF ALGEBRAS IN FIELD THEORY
                         AND PARTICLE PHYSICS "
         Torino, Villa Gualino, September 20 - 30, 1999}
\author{Tom\'{a}\v{s} Kopf${ }^{2}$}\thanks{${ }^{2}$Humboldt Research Fellow}

\address{ThEP, Institut f\"{u}r Physik, Johannes Gutenberg-Universit\"{a}t, \\
55099 Mainz, Germany}

\address{Matematical Institute of the Silesian University at Opava\\
Bezru\v{c}ovo n\'{a}m\v{e}st\'{i} 13 \\ 746 01 Opava, Czech Republic}

\address{Department of Physics, University of Alberta\\
Edmonton T6G 2G1, Canada}

\begin{abstract}
Spacetime, understood as a globally hyperbolic manifold, may be
characterized by spectral data using a 3+1 splitting into space
and time, a description of space by spectral triples and by
employing causal relationships, as proposed earlier.
Here, it is proposed to use the
Hadamard condition of quantum field theory as a smoothness principle.
\end{abstract}

\maketitle

\section{Introduction}

Classical spacetime is to appear from a quantum theory.
Though it is not clear
at present how this is to come about, there is some possibility
that the spacetime manifold will be offered by nature not in the
form of a definition used by some textbook but rather in the
form of spectral data which appear naturally in quantum theory.
Here, such a spectral description of spacetime is discussed.

In A. Connes' noncommutative geometry
\cite{Connes94,Connes95,Connes96a}, a spin manifold is described
using a spectral triple. However, due to the indefinite metric of
spacetime, this scheme is not directly applicable. The problem can
be circumvented by a Hamiltonian description in which spacetime is
foliated by spacelike hypersurfaces, separately describable by
spectral triples and related to each other in a certain way
\cite{Hawkins97}. Such spectral data can be considerably
compressed if causal relationships are exploited \cite{Kopf98}.
This is reviewed in Section \ref{spectraldata}.

In Section \ref{hadamard}, the idea that the spectral data
originate from a quantum theory is taken seriously. The Hadamard
condition of quantum field theory in curved spacetime
\cite{Wald1978a,Wald1978b,Fulling-Sweeny-Wald,
Fulling-Narcowich-Wald,Kay-Wald,Verch1994,Radzikowski1996a,
Radzikowski1996b,Junker,Wellmann,Hollands1999a,Hollands1999b}
is reviewed and proposed as a possible principle
to ensure smoothness.

The conclusion summarizes the presented view and contains some
speculations on how the spectral data as a whole may be generated.

\section{Spacetime spectral data}\label{spectraldata}

A spin manifold with positive definite metric can be described
\cite{Connes94,Connes95,Connes96a} by a certain spectral triple
$(A, \mathcal{H}, D, J,\gamma)$.
 Here, $A$ is a commutative pre-${C}^{\ast}$-algebra
represented (faithfully) on a Hilbert space $\mathcal{H}$, $D$ is
an unbounded selfadjoint operator on $\mathcal{H}$, $J$ is an
antiunitary conjugation and $\gamma$ is a grading operator on
$\mathcal{H}$. These structures satisfy a well known set of conditions
given in \cite{Connes96a}

\begin{note}
The above spectral description is chosen in such a way so as to
make sense in rather general situations, also in the case when the
algebra $A$ is not commutative. In this work which is limited to
classical spacetime with a very simple particle content this
generalization will not be used directly but only as an indication
that the used framework is mathematically a natural one.

In the special setting considered here the algebra $A$ is to be
the algebra of smooth functions ${C}^{\infty}(M)$ on the spin
manifold $M$, $\mathcal{H}$ is the Hilbert space ${L}^{2}(M,S)$ of
sections of the spin bundle, $D$ is the Dirac operator $\not\!\!
D$, $J$ is the charge conjugation and $\gamma$ is the volume
element.
\end{note}

From the spectral triple, it is possible to construct the full
geometry of the spin manifold with positive definite metric. In
particular, the distance between points $p$ and $q$ on the
manifold can be given as the maximal value difference in points
$p$, $q$ for functions with derivative at most 1:

\begin{align}\label{distance}
  d(p,q) &= \sup{\{ \mid f(p)-f(q)\mid ;
  \parallel [\not\!\! D,f]\parallel \leq 1\} }
\end{align}

The assumption of the definiteness of the metric in the above
description is essential. A simple indication of that is visible
in the breakdown of Equation (\ref{distance}): The distance
between two arbitrary points on a manifold with Lorentzian
signature cannot be given by a general length-extremum principle.

While the whole geometry of spacetime with Lorentzian signature
cannot be dealt with by simply taking over the results from
Euclidean signature, it is still possible to describe spacelike
hypersurfaces by spectral triples. The whole spacetime can then be
foliated by such spacelike hypersurfaces ${\Sigma}_{t}$ and by
supplying a time evolution, a lapse function $N$ and shift vector
field ${N}^{i}$, as in the ADM (Arnowitt-Deser-Misner)
formalism \cite{MTW},
the whole manifold can be described in spectral geometry
\cite{Hawkins97}.

In order to avoid inessential technical problems, it will be
assumed that the spacetime manifold $M$ is globally hyperbolic,
i.e., topologically of the form $\Sigma\times\mathbb{R}$ with
$\Sigma$ a compact (spin structure allowing) space manifold and with
the imbeddings ${\iota}_{t}: \Sigma\times\{ t\} \rightarrow
{\Sigma}_{t}\subset M$ providing a foliation by spacelike Cauchy
hypersurfaces.

Thus, at each time, there is a spectral triple $({A}_{t},
{\mathcal{H}}_{t}, {D}_{t}, {J}_{t},{\gamma}_{t} )$ describing $
{\Sigma}_{t}$ which together with the lapse $N$, the shift
${N}^{i}$ and algebra isomorphisms
${\iota}_{{t}_{2}}{\iota}^{-1}_{{t}_{1}}:{A}_{{t}_{1}}\rightarrow
{A}_{{t}_{2}}$ identifying hypersurfaces at different times.

The Hilbert spaces ${\mathcal{H}_{t}}$ can be identified if they
are understood as initial data of a spinor field $\psi$ with mass
$m$ obeying the Dirac equation
\begin{align}\label{diraceq}
  (\not\!\! D -m)\psi = 0
\end{align}
and equipped with the invariant inner product
\begin{align}
  ({\psi}_{1},{\psi}_{2}) = \int_{{\Sigma}_{t}}
  {\bar{{\psi}_{1}}{\gamma}^{\mu}{\psi}_{2}{d\Sigma}_{\mu}}
\end{align}
with ${d\Sigma}_{\mu}$ being the volume element on the
hypersurface and with ${\gamma}^{\mu}$ being the Clifford
generators. Thus, spinor fields given on different spacelike
hypersurfaces are identified, if they are the restrictions of the
same solution of the Dirac equation and the given inner product
does not depend on the hypersurface on which it is calculated.

This Hilbert space has then the meaning of the classical phase
space of the spinor field $\psi$. On this Hilbert space, the
mutually isomorphic algebras ${A}_{t}$ of smooth functions at
different times $t$ are represented differently, allowing to
compare spinor fields localized at different times and thus to
examine causal contacts between hypersurfaces at different times.
It is the examination of the causal contacts between hypersurfaces
that allows drastically to reduce the required spectral data
\cite{Kopf98}.

The knowledge of causal contacts allows in particular to
examine the intersection of the light cone originating at a point
$x\in {\Sigma}_{{t}_{1}}$ with a hypersurface ${\Sigma}_{{t}_{2}}$
at a later time. The light cone intersects the hypersurface
${\Sigma}_{{t}_{2}}$ in a sphere. To first order in time
difference, the size of the sphere gives information on the
distance between the hypersurfaces ${\Sigma}_{{t}_{1}}$,
${\Sigma}_{{t}_{2}}$ and thus provides the lapse function. The
centre of the sphere lies in the normal direction from the tip of
the light cone and allows thus to identify the shift. Since the
sphere has to first order in time difference a constant radius,
independent of direction, also the metric on the hypersurface can
be recovered from the causal relationships, up to a conformal
factor. The conformal factor can, however, be obtained from the
knowledge of the volume element ${\gamma}_{t}$.  At that point
one can dispose of the metric information in the Dirac operators
${D}_{t}$ and keep only its sign ${F}_{t} = {D}_{t}{\mid
{D}_{t}\mid }^{-1}$ which contains information on spatial
smoothness.

The spacetime interval between two events ${p}_{t}$, ${q}_{t+dt}$
occurring at times $t$, $t+dt$ can be given as

\begin{align}\label{stinterval}
  {ds}^{2} &= {dl}^{2}({p}_{t}, {q}_{t+dt})-{dr}^{2}({p}_{t},dt)  \\
  \intertext{where}
{dl}^{2}({p}_{t}, {q}_{t+dt}) &= {\left(\frac {3} {4\pi}\times
\begin{matrix}\text{volume of ball with centre above ${p}_{t}$}\\
 \text{ and ${q}_{t+dt}$ on its boundary}\end{matrix} \right)}^{\frac
{2}{3}}\\ {dr}^{2}({p}_{t},dt) &= {\left(\frac {3}
{4\pi}\times\begin{matrix}\text{volume of ball given at time
$t+dt$}\\ \text{ by light cone originating in
${p}_{t}$}\end{matrix} \right)}^{\frac {2}{3}}
\end{align}

Thus a spectral description of a spacetime manifold may be given
by a family of algebras ${A}_{t}$ represented on a Hilbert space
$\mathcal{H}$ together with families of sign operators ${F}_{t}$,
volume elements ${\gamma}_{t}$ and antiunitary involutions
${J}_{t}$, as given above\footnote{The here given exposition is
based on things like measuring the volume of spheres. This is in
principle possible but is to be understood to be rather of
conceptual value than of practical use since it is not easy to do
calculations in this way. Explicit expressions for the spacetime
geometry, i.e., for the spatial metric, the lapse and the shift
can be obtained from considering commutators of functions in
${A}_{t}$ at different times among themselves and with the
Hamiltonian generating the family ${A}_{t}$, as given in
\cite{Kopf-Paschke2000}. These results can then be used for
practical calculations.}.

However, at this point, it is not possible to spell out a full set
of requirements on the spectral data sufficient to reconstruct a
spacetime manifold {\em without} previous assurance that the data
really give a manifold. In particular, there is no insurance that
the foliated spacetime to be obtained from the data will be smooth
in the time direction. It is the content of the next section to
propose such a global smoothness
principle.

\section{The Hadamard condition}\label{hadamard}

The spectral data in the previous section were lacking a
smoothness principle. However, if the idea is taken seriously
that the phase space of the spinor field $\psi$ which plays the
role of the representation space $\mathcal{H}$ of the spectral
data corresponds to a physical field then such a principle is
naturally present.

In that case, the classical phase space $\mathcal{H}$ generates
via the canonical anticommutation relations a quantum field
algebra and the field is then described by its state on the
algebra. However, not all states on the field algebra are believed
to be physical, since only for very special ones one can define a
sensible stress-energy tensor, as required by R.Wald \cite{Wald1977}
and only for
those states the semi-classical Einstein equations are meaningful.
A sufficient condition for states to be physical in this sense is
the Hadamard condition
\cite{Wald1978a,Wald1978b,Fulling-Sweeny-Wald,Fulling-Narcowich-Wald,
Kay-Wald,Verch1994,Radzikowski1996a,Radzikowski1996b,Junker,Wellmann,
Hollands1999a,Hollands1999b}
 which, after a
reformulation due to Radzikowski
\cite{Radzikowski1996a,Radzikowski1996b}, is understood to be a
positive energy condition on the two-point function of the state.
The technical definition based on micro-local analysis can be
rephrased in the following way:

\begin{definition}\label{hadamarddef}

  The singular structure of the two-point function
  (or of any other distribution)is characterized by its wave front set.
  It is given by all directions in cotangent space
  along which the symbol of any pseudodifferential operator, acting
  on the two-point
  function and producing a smooth result, has to vanish.

  A two-point function is said to satisfy
  the Hadamard condition if its wave
  front set contains only positive
  frequencies propagating forward in time
  and negative frequencies backward in time.
\end{definition}

 States satisfying the
Hadamard condition, so called Hadamard states have the remarkable property
that if they fulfill the Hadamard condition on one Cauchy
hypersurface they do so on the full globally hyperbolic spacetime
as a consequence of its smoothness. It is therefore possible to turn
this argument around and to state the following smoothness
principle:

\begin{smoothness_principle}\label{smoothnessprinciple}
The time evolution given by the spectral data $({A}_{t},
\mathcal{H}, {F}_{t}, {J}_{t},{\gamma}_{t} )$ has to preserve
Hadamard states.
\end{smoothness_principle}

\begin{remark}
The smoothness principle is a necessary condition. It is at this
point not clear to what degree it provides full spacetime
smoothness(see also 1) of the Conclusion), though, in examples, it
can be shown to rule out non-smoothness in the time direction.
\end{remark}
\begin{remark}

  The two-point function is fully determined by operators on the
  classical phase space. This can be done in a particularly simple
  way if the two-point function stems from a quasi-free pure state
  $\omega$ in which case one only needs a complex structure
  ${J}_{\omega}$. It is thus sufficient to work with the classical
  phase space of the spinor field to verify the Hadamard
  condition, a shadow of the full quantum theory. Since the
  Heisenberg picture is used, the Hadamard state $\omega$ (and thus
  ${J}_{\omega}$)is chosen once for all times and is in particular
  independent of the foliation of the considered spacetime. In
  this sense, the smoothness principle is a global one.
\end{remark}

\begin{remark}
  There is a wider class of states, the adiabatic states
  \cite{Parker1969,Deutsch-Najmi,Lueders-Roberts1990,Junker,
  Wellmann,Hollands1999a,Hollands1999b}
  which
  satisfy a weaker, generalized Hadamard condition. They reside  in
  the same folio as the Hadamard states \cite{Junkerprivate}
  \footnote{ see also \cite{Junker-Schrohe}} and
  are easier to construct and to work with. Unfortunately, it is
  not known yet whether they allow for a regularized stress-energy
  tensor but they are a potential alternative to the Hadamard states,
  providing thus an alternative to the smoothness principle.
\end{remark}

\section{Conclusion}

Spacetime, a globally hyperbolic manifold  can be described by
spectral data $({A}_{t},\mathcal{H},{F}_{t},{\gamma}_{t},{J}_{t})$
of Section \ref{spectraldata} and supplied with the smoothness
principle of Section\ref{hadamard}.

However, the spectral data are not to be understood to be in their
final form. A number of open problems and speculations remain and are
to be sorted out in further investigations:

\begin{enumerate}
\item{The smoothness principle may render the data ${F}_{t}$ describing
spatial smoothness superfluous but this is not clear at the
moment. }
\item{The spectral data contain not only information on spacetime
but also preselect a foliation of spacetime. To avoid this, one
may attempt to pack the algebras ${A}_{t}$ into one greater
algebra.}
\item{The algebras ${A}_{t}$ are extremely important in cutting the
classical spacetime out
of the operator algebra on $\mathcal{H}$
but were not given a physical interpretation in the same way as
was done for the spinor field. They are the structures that
provide the classical meaning of position. One possibility is to
interpret them as the shadow of another quantum field interacting
with the spinors field. In particular, if some particles of that
other quantum field would have zero rest mass, it would be
understandable that they would be a primary decoherence inducing
environment \cite{Joos-Zeh,Zurek}, producing thus a pointer basis,
the eigenbasis of ${A}_{t}$. A minimal model would take as the
field in question the $U(1)$ gauge field inherently associated
with a spinor field. The physical content of this theory would then
be quantum electrodynamics on a curved background with the electrons
localized by a bath of photons.}
\end{enumerate}

\section{Acknowledgements}
The author would like to thank Florian Scheck for his hospitality
in Mainz as well as Wolfgang Junker, Don N.Page and Mario Paschke
for a number of useful discussions and Jos\'e M. Gracia-Bond\'{\i}a for
some remarks on a draft version. The support of the Alexander von
Humboldt Foundation as well as previous support from grant VS
96003 "Global Analysis" of the Czech Ministry of Education, Youth
and Sports and from a the NSERC of Canada is most gratefully
acknowledgedged.


\end{document}